# Uncertainty and the Value of Information in Risk Prediction Modeling

Mohsen Sadatsafavi[1], Tae Yoon Lee[1], Paul Gustafson[2]

**Affiliations:**

1. Respiratory Evaluation Sciences Program, Collaboration for Outcomes Research and Evaluation, Faculty of Pharmaceutical Sciences, The University of British Columbia, Vancouver, Canada
2. Department of Statistics, The University of British Columbia, Vancouver, Canada

**Corresponding Author:**

>Mohsen Sadatsafavi, PhD
>Associate Professor, Faculty of Pharmaceutical Sciences
>Centre for Heart Lung Innovation & Dept of Medicine (Respirology)
>University of British Columbia
>CIHR New Investigator, MSFHR Scholar
>http://resp.core.ubc.ca/team/msafavi
>msafavi@mail.ubc.ca

**Short title:** Value Information and Risk Prediction

**Word count:** Abstract: 252; Main document: 4,958 (excluding abstract, appendix, references, acknowledgment, figure legends, and tables)

**Declaration of Conflicting Interests:** None of the authors has any real or perceived conflict of interest with regard to any aspects of this study.

**Financial support:** Financial support for this study was provided in part by a catalyst grant from The Canadian Institutes of Health Research. The funding agreement ensured the authors' independence in designing the study, interpreting the data, writing, and publishing the report.

**Ethics statement:** Ethics approval was not required as the empirical component of this study was based on anonymized, publicly available data (the GUSTO-I trial).

**ABSTRACT**

**Background:** Due to the finite size of the development sample, predicted probabilities from a risk prediction model are inevitably uncertain. We apply Value of Information methodology to evaluate the decision-theoretic implications of prediction uncertainty.

**Methods:** Adopting a Bayesian perspective, we extend the definition of the Expected Value of Perfect Information (EVPI) from decision analysis to net benefit calculations in risk prediction. In the context of model development, EVPI is the expected gain in net benefit by using the correct predictions as opposed to predictions from a proposed model. We suggest bootstrap methods for sampling from the posterior distribution of predictions for EVPI calculation using Monte Carlo simulations. In a case study, we used subsets of data of various sizes from a clinical trial for predicting mortality after myocardial infarction to show how EVPI changes with sample size.

**Results:** With a sample size of 1,000 and at the pre-specified threshold of 2% on predicted risks, the gain in net benefit by using the proposed and the correct models were 0.0006 and 0.0011, respectively, resulting in an EVPI of 0.0005 and a relative EVPI of 87%. EVPI was zero only at unrealistically high thresholds (>85%). As expected, EVPI declined with larger samples. We summarize an algorithm for incorporating EVPI calculations into the commonly used bootstrap method for optimism correction.

**Conclusion:** Value of Information methods can be applied to explore decision-theoretic consequences of uncertainty in risk prediction and can complement inferential methods when developing risk prediction models. R code for implementing this method is provided.

**Highlights:**

- Uncertainty in the outputs of clinical prediction models has largely been approached from a purely statistical perspective.
- In decision theory, uncertainty is associated with loss of benefit because it can prevent one from identifying the most beneficial decision.
- This paper extends Value of Information methods from decision theory to risk prediction and quantifies the expected loss in net benefit due to uncertainty in predicted risks.
- Value of Information methods can complement statistical approaches when developing or validating clinical prediction models.

**INTRODUCTION**

A risk prediction model can be seen as a mathematical function that maps an individual's characteristics to their predicted risk of an event, enabling risk-stratified treatment decisions. The development of a risk prediction model is typically based on individual-level data from a finite sample. As such, the resulting predictions are inherently uncertain. In practice, uncertainty in predictions is often ignored, and a deterministic function is advertised as the final model. For example, the proposed model can be the set of (penalized) maximum likelihood estimates of coefficients in a classical regression framework, or the final state of a machine-learning model such as an artificial neural network. Such determinism in predictions might have stemmed from the need to use the model at point of care where it is most practical to make decisions based on a single good estimate of risk. Notwithstanding such practicality, uncertainty in predictions remains relevant: had we used another sample for model development, we could have arrived at a different model, a different predicted value for the patient, and thus potentially a different treatment decision.

The topic of the development sample size in risk prediction is a subject of active research. Recent developments on sample size calculations have focused on meeting pre-specified criteria on prediction error(1) or on overall calibration performance such as mean calibration or the degree of optimism in predictions(2,3). The adequacy of development sample of a given size has also been investigated in terms of 'stability' of predictions(4). Despite targeting different objectives, such approaches are fundamentally concerned with the accuracy of predictions from a purely statistical perspective. Given that risk prediction models are used for patient care, of ultimate relevance is to what extent such uncertainty affects the outcome of treatment decisions. This perspective to prediction uncertainty is not sufficiently investigated.

We are motivated by the approach taken in the field of decision analysis to tackle a similar problem. In informing policy decisions about the adoption of new interventions, decision-analytic (e.g., cost-effectiveness) models are developed that quantify the net benefit of each competing intervention at the population level(5). Such models are based on uncertain input parameters such as treatment effect or costs of disease management. Thus, the resulting net benefit projections are uncertain. The impact of such uncertainty is that the intervention that is identified as having the highest expected net benefit might not be the one with the highest true net benefit. Consequently, uncertainty is associated with an expected loss in net benefit. The expected value of this loss, termed the Expected Value of Perfect Information (EVPI), can be quantified from the output of a probabilistic decision-analytic model(6). This approach and its extensions, broadly referred to as Value of Information analysis(7), provide a fully decision-theoretic framework for quantifying the impact of uncertainty in health policymaking(8).

In this work, we extend the definition of EVPI from decision analysis to the development phase of risk prediction models, with the aim of quantifying the expected loss in net benefit due to uncertainty in estimation of model parameters from a finite development sample. This provides a decision-theoretic approach to the question that naturally arises after the development of a risk prediction model: whether the model is 'good enough' and can advance to the next stage of research, whether it should be abandoned, or whether more evidence is needed to decide(9,10). The rest of this paper is structured as follows: after reviewing the concept of net benefit in risk prediction, we introduce a Bayesian approach towards net benefit calculations which forms the basis for EVPI definition. We propose a bootstrap-based approach for EVPI calculations. A case study on predicting mortality after myocardial infarction puts the developments in a practical context. We conclude by providing general guidance for interpreting EVPI, discussing the relevance of other sources of uncertainty, reviewing potential applications of Value of Information in other aspects of clinical prediction modeling, and suggesting areas of further research.

**Net benefit calculations for risk prediction models**

The net benefit approach for evaluating the utility of risk prediction models has gained significant popularity due to its rigorous decision-theoretic underpinning compared with purely statistical approaches such as calibration or discrimination, as well as its relative ease of calculation(11). In order to turn a continuous predicted risk to a binary action (treat or not treat), one needs to specify a context-dependent treatment threshold on predicted risks. Such a threshold should ideally be informed by the relative weight of clinical consequences of false-positive (harm) versus true-positive (benefit) classifications. Vickers and Elkin showed that this threshold acts as an exchange rate between true and false positive outcomes, enabling the calculation of net benefit(10,11). Imagine, for example, a decision-maker (e.g., a clinical guideline development team, after consulting a patient group about their preferences) concludes that patients with acute myocardial infarction (AMI) should receive a more aggressive treatment if their 30-day risk of mortality is >2%, and no such treatments if the predicted risk is <2%. The group is ambivalent between treatment and no treatment if the predicted risk is precisely 2%. Such ambivalence indicates that the decision-maker equates the benefits associated with 2% chance of true positive to be equal to the harms associated with a 98% chance of false positive. This itself means the benefit of a true positive diagnosis is 49 times the harm of a false positive diagnosis. This enables the calculation of net benefit (NB) in true positive units net of harms in false positive units . Generalizing this approach, at threshold value of $z$, the NB can be calculated as

$$NB(z) = P(True\ Positive) - P(False\ Positive)\frac{z}{1-z}.$$

Here, $z/(1-z)$ represents the relative weight of a false positive versus a true positive classification and thus captures the harm-benefit trade-off associated with this threshold. In practice, the NB is often calculated for a plausible range of thresholds.

Imagine we have a 'proposed' model based on a development sample of $n$ independent observations. Let $\pi_i \equiv \pi(X_i)$ be the predicted risks for the i[th] patient in this sample with covariate pattern $X_i$, and $Y_i$ be the corresponding observed binary outcome. At threshold value of $z$, the i[th] patient contributes $I(\pi_i > z)Y_i$ to the probability of true positive and $I(\pi_i > z)(1-Y_i)$ to the probability of false positive. The NB of the proposed model can thus be consistently estimated as(11)

$$\widehat{NB}_{model}(z) = \frac{1}{n}\sum_{i=1}^{n}\left\{I(\pi_i > z)\left[Y_i - (1-Y_i)\frac{z}{1-z}\right]\right\}.$$

The NB of the model should always be compared with that of at least two alternatives: treating none and treating all. We use the 'opt-in' definition of NB and set the default decision to be treating no one, with NB=0(9). The decision to treat all is equal to assuming each individual is considered positive, whose NB can be consistently estimated as

$$\widehat{NB}_{all}(z) = \frac{1}{n}\sum_{i=1}^{n}\left\{Y_i - (1-Y_i)\frac{z}{1-z}\right\}.$$

If there are pre-existing models that are applicable to this decision context, their NB should also be compared with the NB of the model. However, to facilitate the developments and without loss of generality, we assume the proposed model is the only relevant risk prediction algorithm.

Evaluating a model in the same sample in which it is developed might result in optimistic conclusions about its performance(12). A commonly employed method for correcting for such optimism is the Harrel's bootstrap(13). This approach involves obtaining a bootstrap sample from the development dataset, fitting a new model in this sample, and calculating NB (or other metrics of interest) for the new model in the same bootstrap sample as well as in the original sample, and recording the difference. Repeating these steps many times and averaging the results will provide an estimate of optimism. This approach is based on the notion that the difference between the performance of the model in the bootstrap sample and in the original sample is an almost unbiased estimate of the difference between its performance in the original sample and in the generating population(14).

**A Bayesian approach towards net benefit calculation**

Value of Information analysis is a strictly Bayesian paradigm as it treats the unknown true associations as random entities for which the development sample provides partial information(6). Here, the random entity of interest is the 'correct' model, indexed by a set of unknown parameters $\theta$, that for the i[th] individual returns the correct risk $p_{\theta i} \equiv p(X_i, \theta)$, the average risk among all individuals with the same covariate pattern $X_i$. Let $P(\theta|D)$ be the posterior joint probability density function of model parameters that represents our knowledge about the parameter values of the correct model after observing the development data $D$. The Bayes' rule $P(\theta|D) \propto P(\theta)P(D|\theta)$ indicates that our knowledge is influenced by the information from the development sample ($P(D|\theta)$) and any prior knowledge on the correct model ($P(\theta)$).

The crucial next step is to recognize that if the correct risks are available, we can replace the observed response $Y_i$ with the correct risk $p_{\theta i}$ for estimating the NB of the proposed model. At threshold value of $z$, the i$^{th}$ person with a predicted risk of $\pi_i$ and correct risk of $p_{\theta i}$ has, on average, a probability of $I(\pi_i > z)p_{\theta i}$ for being a true positive and $I(\pi_i > z)(1 - p_{\theta i})$ for being a false positive. Thus, if the true value of $\theta$ is known, we can consistently estimate the NB of the model as

$$NB_{model}(z; \theta) = \frac{1}{n}\sum_{i=1}^{n}\left\{I(\pi_i > z)\left[p_{\theta i} - (1 - p_{\theta i})\frac{z}{1-z}\right]\right\}.$$

This equation is similar to the equation for $\widehat{NB}_{model}$, only that the $Y$ column in the development sample is replaced with predicted risks from the correct model. As we do not know the true value of $\theta$, in our Bayesian framework estimating NB at threshold $z$ requires taking the expectation with respect to $P(\theta|D)$:

$$\overline{NB}_{model}(z) = \mathrm{E}_\theta NB_{model}(z; \theta).$$

Unlike the conventional estimator for NB, this estimator is the posterior mean in a Bayesian framework and the frequentist notion of optimism is not directly applicable to it: rather than being based on a single value of $\theta$ that might provide an overly good fit to the data, it is the average of NB estimates across the distribution $P(\theta|D)$. Again, using the risk prediction model is not the only option as we can also either forgo treatment for all or provide treatment to all. The former has zero NB, and the NB for the latter, if the true value of $\theta$ is known, can be estimated consistently as

$$NB_{all}(z; \theta) = \frac{1}{n}\sum_{i=1}^{n}\left[p_{\theta i} - (1 - p_{\theta i})\frac{z}{1-z}\right],$$

which is again the same as $\widehat{NB}_{all}$, with the $Y$ column replaced by correct risks. The expected net benefit of treating all with respect to the distribution of $\theta$ is

$$\overline{NB}_{all}(z) = \mathrm{E}_\theta NB_{all}(z; \theta).$$

**The Expected Value of Perfect Information (EVPI)**

If we know the correct model, the optimal decision is to use it, instead of the proposed model, for prediction. Indeed, no decision that is based on candidate predictors is more efficient than giving treatment only to those whose correct risk, based on such predictors, is above the threshold. If the true $\theta$ is known, the NB of such an optimal strategy can be estimated consistently in the sample as

$$NB_{max}(z;\theta) = \frac{1}{n}\sum_{i=1}^{n} I(p_{\theta i} > z)\left[p_{\theta i} - (1 - p_{\theta i})\frac{z}{1-z}\right].$$

Again, we do not know the true value of $\theta$, and instead know about its likely values through $P(\theta|D)$. The expected NB under perfect information is therefore

$$\overline{NB}_{max}(z) = \mathrm{E}_\theta NB_{max}(z;\theta).$$

On the other hand, without knowing the correct model, the best we can do is to decide whether to use the model, treat no one, or treat all, at the given threshold, based on their expected NB. The expected NB under current information is therefore $\max\{0, \overline{NB}_{model}(z), \overline{NB}_{all}(z)\}$.

The difference in expected NB with perfect information compared with current information is the expected gain due to knowing the correct model (or expected loss due to not knowing the correct model), which is called the Expected Value of Perfect Information (EVPI) (15):

$$EVPI(z) = \overline{NB}_{max}(z) - \max\{0, \overline{NB}_{model}(z), \overline{NB}_{all}(z)\}.$$

EVPI is a non-negative scalar quantity that is in the same unit as the NB for risk models, and its higher values indicate higher expected loss due to prediction uncertainty.

**Relative EVPI**

The scale of NB in risk prediction is domain-specific, unlike in decision analysis where NB and thus EVPI are typically in the universally interpretable monetary units. As such, the numerical value of EVPI (i.e., the magnitude of expected NB loss due to uncertainty) is the most interpretable in comparison with the expected NB that the model provides. To facilitate this comparison, we suggest a relative version of EVPI. To proceed, we note that without using any model, we can choose between treating none or treating all. At a given threshold, this strategy confers an expected NB of $\max\{0, \overline{NB}_{all}(z)\}$. This is the 'baseline' benefit without any risk stratification. Against this baseline, the expected incremental NB (ΔNB) of using the proposed model is

$$\overline{\Delta NB}_{current\ information}(z) = \max\{0, \overline{NB}_{model}(z), \overline{NB}_{all}(z)\} - \max\{0, \overline{NB}_{all}(z)\}.$$

Similarly, the expected incremental NB with knowing the correct risks is

$$\overline{\Delta NB}_{perfect\ information}(z) = \overline{NB}_{max}(z) - \max\{0, \overline{NB}_{all}(z)\}.$$

The EVPI is the difference between the two terms. We suggest 'relative EVPI' (EVPI$_r$) as the ratio of the two terms:

$$EVPI_r(z) = \frac{\overline{\Delta NB}_{perfect\ information}(z)}{\overline{\Delta NB}_{current\ information}(z)}.$$

This quantity is ≥1 (since $\overline{\Delta NB}_{perfect\ information}(z) \geq \overline{\Delta NB}_{current\ information}(z) \geq 0$) and can be expressed in percentages. An $EVPI_r$ of 1+$\alpha$ means that against the baseline strategy of not using any model, the expected gain in NB with the use of the correct model is $\alpha$*100% higher than the expected gain in the NB with the use of the proposed model. The $EVPI_r$ is +∞ when the denominator is zero but the numerator is positive. This indicates that under current information, the proposed model is not expected to provide extra NB, but the correct model is. Thus, further development attempts might be justified. The $EVPI_r$ is undefined when the numerator (and thus the denominator) is zero but the conclusion is obvious in this case: the correct model, and therefore the proposed model, are not expected to provide extra NB over the default decisions, regardless of current uncertainties.

**Risk of decision reversal**

The value of EVPI is affected by both the probability of incorrectly identifying the best decision (the probability of NB loss) and the consequence of such an incorrect identification (the magnitude of NB loss). In the proposed Bayesian framework, the probability of incorrect identification of the best decision, which we call the *risk of decision reversal*, can be calculated and reported separately as a useful auxiliary metric. For example, if using the correct model is the current best decision, then

$$P(NB_{model}(z;\theta) < \max\{0,\ NB_{all}(z;\theta)\}) = \mathrm{E}_\theta I(NB_{model}(z;\theta) < \max\{0,\ NB_{all}(z;\theta)\})$$

is the risk of decision reversal at threshold value of $z$ (with I() being the indicator function).

**A generic algorithm for EVPI calculation based on bootstrapping**

The Bayesian estimators in the previous sections require taking expectations with respect to $P(\theta|D)$, the posterior distribution of correct model parameters. A fully Bayesian approach to developing a risk prediction model enables the specification of $P(\theta|D)$ given the development data and any prior information. Alternatively, in the conventional likelihood maximization approach in classical regression modeling, $P(\theta|D)$ can be derived from the likelihood function. For example, the vector of maximum likelihood estimates of regression coefficients and their estimated covariance matrix from a logistic regression model can be used to specify a multivariate normal distribution as the posterior distribution of regression coefficients under an improper, flat prior. The expectations can then be evaluated using Monte Carlo simulation with repeated sampling from $P(\theta|D)$.

A more flexible approach is to obtain samples from $P(\theta|D)$ via bootstrapping. A Bayesian interpretation of the bootstrap enables one to consider a parameter estimate that is derived from a bootstrapped sample as a random draw from the posterior distribution of the parameter given the original sample(16). A Bayesian bootstrap of a sample of $n$ i.i.d. observations is obtained by drawing a random vector of weights $(w_1, w_2, \dots, w_n)$ from a $Dirichlet(n; 1,1, \dots ,1)$ distribution; i.e., we obtain a weighted sample, with the total sum of weights being $n$(16). This can be motivated as a (nonparametric) posterior draw for the data-generating distribution, under a non-informative (and improper) prior distribution. One way to generate such weights is drawing $n-1$ standard uniform random variables $u_1, \ldots, u_{n-1}$, ordering them, and calculating the weights as $w_i = u_i - u_{i-1}$, where $u_0 = 0$ and $u_n = 1$(16).

The ordinary bootstrap can also be seen as assigning weights to the sample, with weights coming from $Multinomial(n; 1/n, \ldots, 1/n)$. The similarity of such weighting approach has resulted in the ordinary bootstrap being also interpreted in a Bayesian view. For example, the approximate Bayesian bootstrap method for the imputation of missing data is based on interpreting the ordinary bootstrap as a Bayesian one(17). Such a bootstrap-based Value of Information approach for data-driven decision analysis has been previously proposed, and it is shown that Bayesian and ordinary bootstraps and parametric methods generate very similar results(18).

This bootstrap-based approach for sampling from $P(\theta|D)$ provides more flexibility than fully parametric methods, for example by enabling the incorporation of variable selection and shrinkage (thus incorporating model uncertainty in estimations), and potentially other stochastic steps such as the imputation of missing predictor values. As well, this approach can be embedded with relative ease within the bootstrap-based algorithm for optimism correction. A generic algorithm for EVPI calculation can be formulated as follows:

- Using the proposed prediction model, generate the predicted risks for each individual in the development sample ($\pi_i$s).
- For i=1 to some large N (e.g., 1,000)* and for any threshold z in [0,1]
  - Obtain a (Bayesian) bootstrap sample from the development dataset, and perform model development (including variable selection and shrinkage).
  - Apply the new model to calculate the predicted risks in the development sample ($p_{\theta i}$s).
  - Estimate $NB_{model}(z)$, $NB_{all}(z)$, and $NB_{max}(z)$ for this iteration using the predicted risks from the original ($\pi_i$s) and the new ($p_{\theta i}$s) model in the development sample (see the relevant equations in the text).
- Let $\overline{NB}_{model}(z) = $ average$(NB_{model}(z))$; let $\overline{NB}_{all}(z) = $ average$(NB_{all}(z))$; let $\overline{NB}_{max}(z) = $ average$(NB_{max}(z))$, with the average taken across all the iterations from the For loop.
- Calculate (absolute and relative) $EVPI(z)$.

*In general, the number of iterations should be high enough such that the Monte Carlo standard error around EVPI is small compared to its point estimate.*

**Case study: prediction of mortality after acute myocardial infarction (AMI)**

Identifying the risk of immediate mortality after an AMI can enable stratification of more aggressive treatments for high-risk individuals. GUSTO-I was a large clinical trial of multiple thrombolytic strategies for AMI(19). We used data from this study to create a risk prediction model for 30-day mortality after AMI (the primary end-point of the trial). GUSTO-I's sample size of 40,830 is larger than typical sizes of development samples in most practical contexts – at least outside of cardiovascular diseases, resulting in low level of prediction uncertainty if the full sample is used(20). This provides an opportunity for simulating development samples of smaller sizes that are more typical, and studying how EVPI changes as the sample size varies. To start, we assume that we have access to data for only 1,000 patients. We randomly selected, without replacement, 1,000 individuals from the full sample of GUSTO-I to create such an exemplary development dataset. 30-day mortality risk was 7.0% in the full sample and 6.7% in this subsample.

In line with previous studies using this dataset(21,22), our candidate predictors included Killip score (an indicator of heart failure), age, blood pressure, pulse, location of the infarction within the heart, as well pre-existing hypertension and diabetes. To mitigate the risk of overfitting, we fitted a logistic regression model via the least absolute shrinkage and selection operator (LASSO), with 10-fold cross-validation to find the optimum shrinkage, targeting mean squared prediction error. **Table 1** provides the coefficients of the proposed model. Three candidate predictors were shrunk to zero (not selected) in the final model. To demonstrate uncertainty in regression coefficients, we also report the bootstrap-based 95% confidence intervals and the proportion of bootstraps in which each predictor was selected by LASSO. Confidence intervals, optimism-corrections, and EVPI calculations are from Monte Carlo simulations based on 1,000 bootstraps. Computations are performed in R development environment(23). The *glmnet* package was used for fitting the LASSO models. The code and data are available from https://github.com/Shoodood/VoIPred/.

*Table 1:* Regression coefficients for the proposed model

| Predictors | Coefficient* | Probability of selection | 95% confidence interval |
|---|---|---|---|
| (Intercept) | -1.273 | 1.00 | -6.833, 2.744 |
| Age (years) | 0.050 | 1.00 | 0.021, 0.076 |
| AMI location (other) | 0.259 | 0.55 | -0.034, 1.488 |
| AMI location (anterior) | . | 0.40 | -0.220, 0.521 |
| History of previous AMI | 0.0184 | 0.63 | -0.058, 0.842 |
| Systolic blood pressure† | -0.070 | 0.94 | -0.104, 0.000 |
| Killip score >1 (yes v. no) | 0.704 | 0.97 | 0.000, 1.277 |
| Pulse (low) ‡ | 0.026 | 0.75 | 0.000, 0.038 |
| Pulse (high) ‡ | . | 0.40 | 0.000, 0.034 |
| History of hypertension | . | 0.31 | -0.587, 0.259 |
| History of diabetes | . | 0.38 | -0.218, 0.664 |

*Those denoted by '.' are not selected by LASSO.
†This variable was modeled as min(X,100).
‡Pulse was modeled using a linear spline with a knot location at 50.
*AMI: Acute myocardial infarction*

The optimism-corrected c-statistic of the proposed model was 0.758. *Figure 1* is the 'Decision Curve' that depicts the optimism-corrected empirical NB ($\widehat{NB}_{model}$) of the model (red) alongside those of treating none (gray) and treating all (black). The Bayesian estimator for NB ($\overline{NB}_{model}$, blue curve) is also provided. Ordinary and Bayesian bootstraps generated nearly identical results.

*Figure 1:* Optimism-corrected (red) NB of the proposed model and its Bayesian estimator* (blue), compared with the NB of treating all (black) and treating none (gray)#

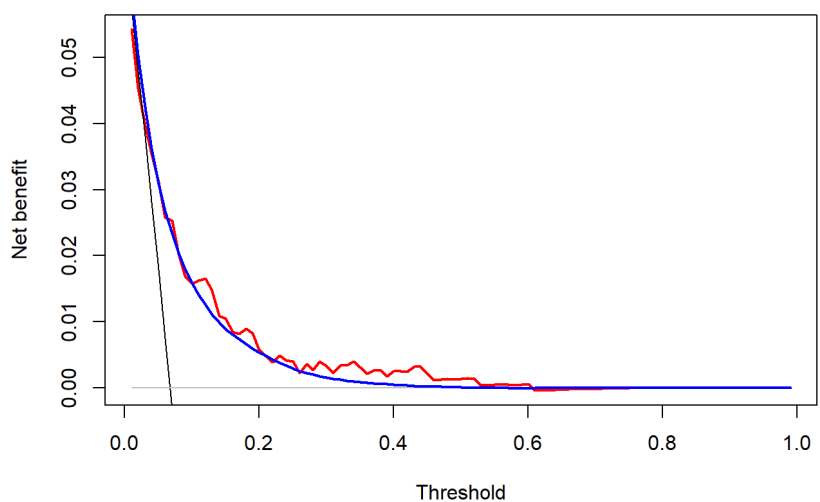

*The Bayesian estimation is based on the Bayesian bootstrap (see the relevant section in the text)
#Optimism-correction and Bayesian estimates are based on 1000 bootstraps.
*NB: net benefit*

*Figure 2* depicts the expected incremental NB under current and perfect information (left panels) and EVPI (right panels) at the entire range of thresholds. Results are generated using both ordinary (approximate Bayesian) and Bayesian bootstraps, which were very similar. Interpreting the results based on the ordinary bootstrap, at the exemplary threshold of 0.02, the expected NB of treating all is 0.0478, while the expected NB of the model is 0.0484. Thus, the best decision under current information is to use the proposed model, with an expected incremental benefit of 0.0006 (black curve in top-left panel). The expected NB under perfect information is 0.0489, corresponding to an expected incremental NB of 0.0011 (red curve in top-left panel). Thus, the EVPI is 0.0489-0.0484=0.0005. The relative EVPI at this threshold is 0.0011/0.0006=1.87. That is, knowing the correct prediction model is expected to confer 87% more NB compared with the proposed model. The risk of decision reversal at this threshold is 27%. The EVPI is non-zero unless the threshold is unrealistically high (>0.85). The largest gain is obtained within the 0.1–0.3 threshold range. The Bayesian bootstrap generated similar results (EVPI at 0.02 threshold: 0.0005, relative EVPI at this threshold: 1.82, risk of decision reversal: 27%).

*Figure 2:* The Incremental Net Benefit curves under perfect (black) and current (red) information (left) and EVPI (right)

Ordinary (approximate Bayesian) bootstrap

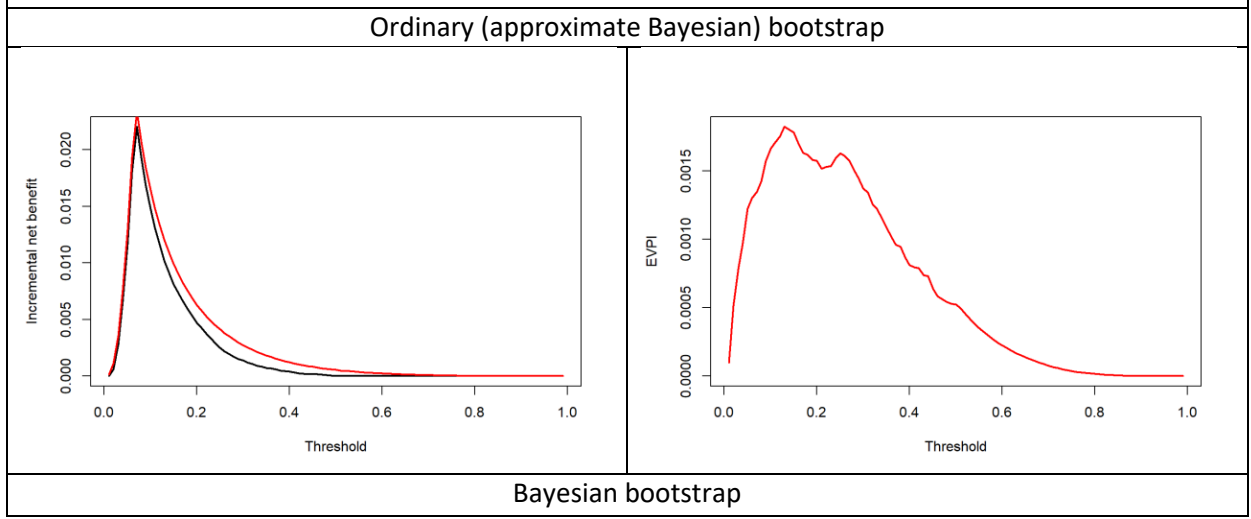

Bayesian bootstrap

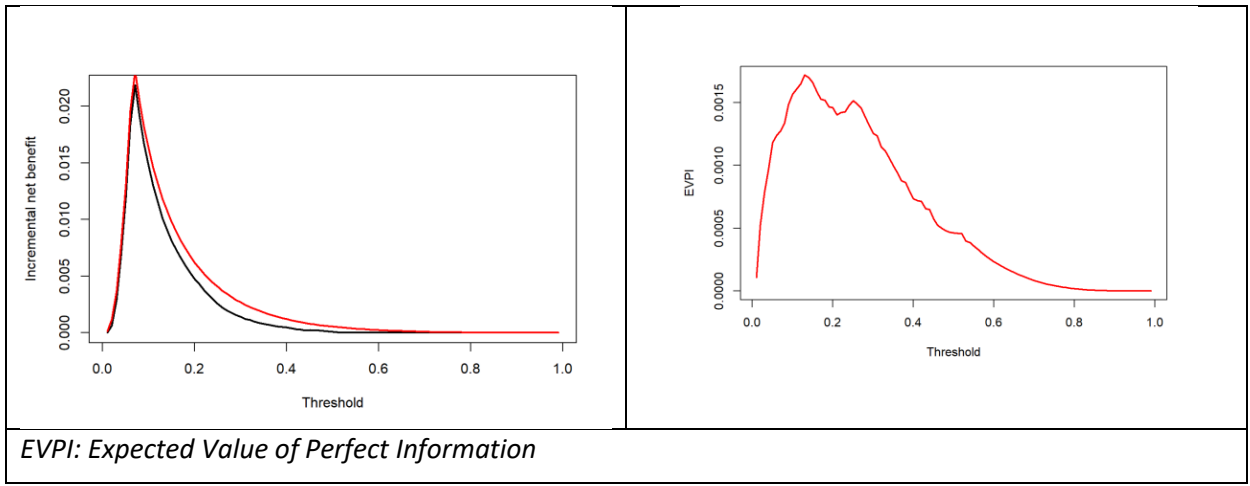

*EVPI: Expected Value of Perfect Information*

*Figure 3* demonstrates how EVPI changes with sample size. We simulated development samples from GUSTO-I, starting from 500 observations and doubling the size at each step. The left ad rights panels demonstrate, respectively, EVPI and relative EVPI, at the exemplary thresholds of 0.01, 0.02, 0.05, and 0.10. Both metrics indicate a diminishing gain with larger samples. For the model based on the entire GUSTO-I data, the impact of uncertainty was minimal, with EVPI being <0.00001 at threshold value of 0.02, and relative EVPI being 1.004.

**Figure 3:** Change in EVPI at threshold value of 0.1 (left) and relative EVPI (right) as a function of sample size*

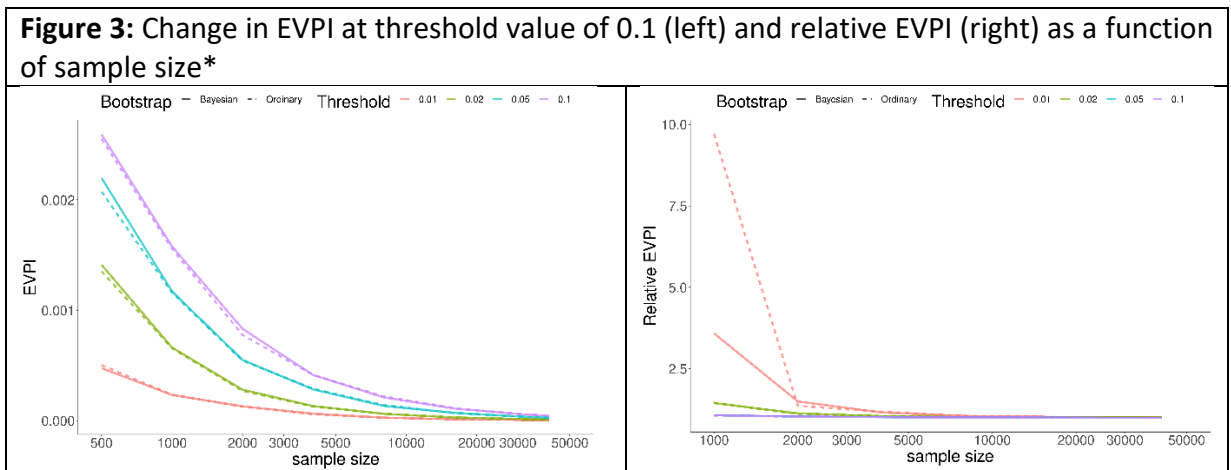

*The largest sample is 40,830

Results were obtained based on randomly obtaining samples, without replacement, of a given size. Results are averaged (left panel) and median (right panel) across 10 independent simulations for each sample size, with EVPI calculations based on the ordinary bootstrap with 1,000 iterations. We discarded datasets with fewer than 8 events as the *glmnet* optimizer does not reliably converge with too few events. For relative EVPI, the smallest

> sample size was 1,000 because more than 50% of values for the sample size of 500 were +∞.
>
> *EVPI: Expected Value of Perfect Information*

**DISCUSSION**

Creating a risk prediction model based on a finite development sample means the resulting predictions are inevitably uncertain. The management plan of a patient based on such predictions might be different from the decision that would have been made had the correct risks been known. As such, prediction uncertainty can result in loss of net benefit. We extended the Value of Information methodology from decision analysis to the development phase of risk prediction models and applied the definition of EVPI to this context. The proposed development EVPI is a scalar metric that quantifies, for a given risk threshold, the expected loss due to uncertain predictions, with the loss being defined on the same net benefit scale as is commonly used to assess the utility of the risk prediction models(11). In a case study using data from a clinical trial, we demonstrated how EVPI can be calculated and interpreted, for example by determining the range of thresholds within which obtaining a larger development sample could potentially be warranted. We also showed how EVPI behaves when the development sample size is increased. We proposed relative EVPI as a scale-free metric, and outlined a generic bootstrap-based algorithm for EVPI calculations that can be embedded within established algorithms for quantifying the optimism of risk prediction models.

How should these developments be used in practice? Once the risk model is developed, the investigators need to decide whether the model is good enough to go to the next stage (i.e., should move from development to validation), should be abandoned, or further model development is required(9). Classical arguments in decision theory stipulate that under the conditions of risk-neutrality and absence of sunk costs (irrecoverable costs associated with implementing a health technology), the 'adoption decision' and 'research decision' are independent: it is solely the expected NB of alternative decisions that should determine whether to adopt the model or not(24), while Value of Information metrics determine whether further evidence (e.g., obtaining a larger development sample) is required. However, model developers as scientists generally have a preference against seeing their discoveries proven incorrect or harmful(25), and patients, care providers, and the general public are on average risk-averse, having a preference towards avoiding harm than causing or accruing benefit(26,27). As well, there are significant sunk costs associated with implementing a risk stratification algorithm only to abandon it later (updating guidelines, incorporating the algorithm into Electronic Health Records). Given these, uncertainty and the resulting potential for harm become relevant factors when deciding whether a model should advance to the next stage.

If risk behaviors in the given clinical domain are to be considered, one can update the decision criterion and Value of Information equations by explicitly considering risk attitudes(28). However, in early phases of model development, investigators might be unwilling to make judgment calls about risk attitudes. We think in this phase what is the most helpful is general guidance on whether the decision-theoretic impact of predictions uncertainty is low enough that justifies moving towards model validation. In this context, a zero EVPI indicates that the currently identified best decision is the correct one in this patient population. Similarly, a low EVPI indicates that the potential for harm with current information is small. Such results can motivate model developers to focus on the next stage (e.g., depending on the NB of the model, abandon the model or move to validation). On the other hand, when EVPI is large, the potential for harm is significant and it might be rational not to proceed before an updated model based on a larger development sample is produced. This invokes the question of what value of EVPI is large enough to warrant pursuing further model development. While this is context-specific, for the development phases it might make sense to specify thresholds on EVPI as general guidance. For example, an EVPI that is similar to the incremental ΔNB of the model (i.e., relative EVPI of ~2) indicates that the expected loss due to uncertainty is equal to the expected gain by using the proposed model. This can be interpreted as the presence of substantial uncertainty and potential for harm. Such a threshold on (relative) EVPI can be more relatable than thresholds on statistical metrics such as calibration slope or degree of optimism, whose implications for medical decisions are less clear. This approach can thus potentially lead to stronger consensus among stakeholders and defendable recommendations by authorities who formulate best practice standards in predictive analytics.

The EVPI as defined in this work represents the uncertainty due to the finite development sample, resulting in uncertainty in the regression coefficients of the prediction model. Importantly, this EVPI does not represent the value of knowing the true risk for each individual, which is also a function of predictors that are unknown, unmeasured, or intentionally left out of the model, rather the correct expected risk among all individuals with the same values of candidate predictors (i.e., perfect information on regression coefficients). It also does not include uncertainty due to the potentially systematic differences between the development and the target population (related to external validation which is discussed below). However, modifications of this definition are conceivable that can bring other sources of uncertainty into consideration. Consider, for example, that there is an otherwise strong predictor of outcome in the development sample that is intentionally excluded due to difficulty in measuring it in practice. If in the Monte Carlo bootstrap algorithm for producing draws from $P(\theta|D)$ one includes this predictor in regression models, the resulting EVPI combines the expected loss due to the finite development sample and due to not including the predictor. Similarly, if there are predictors with missing values, incorporating the process of imputing such missing values within Monte Carlo iterations (i.e., stochastically imputing missing values within each bootstrap set) means that the resulting EVPI represents the loss due to the finite development sample and due to missing data.

The Bayesian inference underlying EVPI calculations is based on the assumption that the prior distribution $P(\theta)$ and the data model $P(D|\theta)$ are compatible with the true data generating mechanism. Under these assumptions, Bayesian posterior distributions are guaranteed to be calibrated (in contrast with the frequentist inference where a correct model structure by no means prevents overfitting)(29). These assumptions are similar to the assumptions that enable Value of Information calculations in decision analysis: that the model structure is correct (such that plugging in the true values of input parameters would result in correct net benefits) and the probability distributions correctly specify our uncertainty about the values of input parameters. It is indeed improbable that these assumptions are fully met in practice, as both decision-analytic and risk prediction models are simplifications of reality. Nonetheless, Value of Information in decision analysis is justified based on the working assumption that a model that is good enough for calculating net benefit is also good enough for quantifying uncertainty around it. We think this assumption is generally a reasonable one in risk modeling. Nevertheless, this framework should be used with caution with black-box algorithms such as Machine Learning models. Given that such models typically have many free parameters, the cost of model misspecification can be particularly high. In general, to what extent Value of Information quantities are robust against departures from correct model specification needs to be studied.

The application of Value of Information to risk prediction can be a fruitful endeavor on many fronts. An important area of inquiry is the application of this concept to external validation of risk prediction models. Unlike during model development when the ultimate goal is to identify the correct model, in external validation the goal is to evaluate if a 'prespecified' model performs well and thus using it will be beneficial. The expected gain by perfectly knowing if a prespecified model is net beneficial in a new population is different from the expected gain by knowing the correct model in this population. As such, the EVPI for external validation is distinct from the development EVPI and needs to be developed independently. Further, the Expected Value of Sample Information is a related metric in decision analysis that quantifies the expected gain in net benefit from conducting a specific study with a given design and sample size(8). Defining the equivalent of this metric for risk prediction seems feasible and an immediate extension of the proposed framework. Net benefit calculations have been extended from risk prediction models to models that aim at predicting the benefit of specific interventions(30) and Value of Information methods can conceivably be extended to such context.

In conclusion, contemporary approaches towards evaluating uncertainty in risk prediction target prediction error, calibration, or stability. Despite significant contributions, these metrics are statistical in nature, as they do not relate prediction uncertainty to the outcome of medical decisions. Evaluating the net benefit of a risk prediction has complemented purely statistical approaches for the assessment of risk prediction models, in a way that is considered a 'breakthrough' in predictive analytics(9). We think the assessment of uncertainty in such models can also be augmented with a decision-theoretic perspective.

**Appendix:** Stylized R code for EVPI calculation. The R code uses an exemplary dataset in the MASS package.

```r
library(MASS)
data_set <- birthwt #An exemplary dataset containing the birthweight status of newborns and some covairates

n <- dim(birthwt)[1]

z <- 0.2 #This is the risk threshold

model <- glm(low ~ age + lwt, family=binomial(link="logit"), data=data_set) #Our risk prediciton model
pi <- predict(model, type="response") #Predicted risks

#We hold simulation results in memory for exploration (not necessary)
out <- data.frame("NB_all"=double(),"NB_model"=double(),"NB_max"=double())

for(i in 1:1000) #Looping over 1000 simulations
{
  bs_data_set <- data_set[sample(1:n, n, replace = T),] #Create a bootstrap sample and fit the model again
  bs_model <- glm(low ~ age + lwt, family=binomial(link="logit"), data=bs_data_set)
  #Predict risks from this model are applied to the original data
  p <- predict(bs_model, newdata = data_set, type="response")

  #Bayesian NB calculations. p is taken as random draws from the distribution of correct risks
  NB_all <- mean(p-(1-p)*z/(1-z)) #NB of treating all
  NB_model <- mean((pi>z)*(p-(1-p)*z/(1-z))) #NB of using the model
  NB_max <- mean((p>z)*(p-(1-p)*z/(1-z))) #NB of using the correct risks

  out[i,] <- c(NB_all,NB_model,NB_max)
}

EVPI <- mean(out$NB_max) - max(0,mean(out$NB_all),mean(out$NB_model))
cat("EVPI is ",EVPI)

#Some additional results
cat("Without any model, the expected NB of the best decisoin is ", NB_base <- max(0,mean(out$NB_all)))
cat("The expected incremental NB of the proposed model is ", INB_current <- 
max(0,mean(out$NB_all),mean(out$NB_model)) - NB_base)
cat("The expected incremental NB of the correct model is ", INB_perfect <- mean(out$NB_max) - NB_base)
cat("Relative EVPI is ", INB_perfect / INB_current)
```